\documentclass[aip,graphicx]{revtex4-1}

\usepackage{graphicx}% Include figure files
\usepackage{dcolumn}% Align table columns on decimal point
\usepackage{bm}% bold math
\usepackage{wasysym}
\usepackage{siunitx}

\draft % marks overfull lines with a black rule on the right

\begin{document}

%\preprint{}

\title{Stimulation of surface ionization waves by pulsed laser irradiation} %Title of paper



\author{Thomas Orrière}
\author{David Z. Pai}
\email[]{david.pai@lpp.polytechnique.fr}
%\thanks{}
\altaffiliation[Current affiliation: ]{Laboratoire de Physique des Plasmas (LPP), CNRS, \'Ecole Polytechnique, Sorbonne Université, 91128, Palaiseau, France}
\affiliation{Institut PPRIME (CNRS UPR 3346, Université de Poitiers, ISAE-ENSMA)\\
		86962, Chasseneuil Futuroscope, France}


\date{\today}

\begin{abstract}
The inclusion of semiconducting material within a composite barrier enables the perfectly uniform propagation of surface ionization waves (SIW) in air at atmospheric pressure regardless of the polarity of the applied electric field, unlike surface discharges generated using purely dielectric barriers. We exploit the photonic properties of silicon to stimulate the SIW using external irradiation by a 2-ns pulsed laser at 532 nm, with a fluence of 1.3 mJ/cm$^2$ per pulse at the surface. No effect is observed when irradiation occurs more than 3 µs before plasma generation. This timescale is attributed to the ambipolar diffusion of photoexcited carriers away from the Si-SiO$_2$ interface. When this delay shortens to less than 3 µs, the SIW propagates farther and with more intense optical emission. Furthermore, the energy of the discharge increases by up to 7\%. The sensitivity to the laser-plasma delay demonstrates that the observed stimulation of the SIW cannot be due to the desorption of surface charge by irradiation.
\end{abstract}


\maketitle %\maketitle must follow title, authors, abstract and \pacs


\section{\label{sec:level1}Introduction}
	
Surface ionization waves (SIW) are fundamental elements of two major classes of atmospheric-pressure low-temperature plasmas: surface dielectric barrier discharges (SDBD) and plasma jets. SDBDs have typically been implemented in planar geometries, notably for applications in electroaerodynamic (EAD) flow control~[\onlinecite{benard2014electrical}] and more recently EAD propulsion~[\onlinecite{xu2019dielectric,gomez2021performance,wilde2021model}] as "emitters" or sources of ions that drive electromechanical energy conversion when drifting through an applied electric field. SDBD are also employed in plasma-assisted ignition and combustion for the production of reactive species and fast gas heating, as well as possible hydrodynamic effects~[\onlinecite{shcherbanev2017ignition}]. Plasma jets are generated from within dielectric tubes, analogous to SDBD rolled into a cylindrical rather than planar geometry. Under certain conditions, the discharge is initially an annular SIW while still propagating inside the dielectric tube, prior to its exit into ambient gas as a plasma jet~[\onlinecite{viegas2022physics}].

In many gases at atmospheric pressure, individual ionization waves typically propagate in the form of a thin channel, commonly referred to as a streamer~[\onlinecite{raizer1997gas}]. The radius of the streamer channel, as well as the radius of curvature of the strongly ionizing head of the streamer, are much smaller than its length. Streamers also occur in groups, as is the case for SDBDs, where SIW can propagate in closely packed form that may be considered quasi-uniform~[\onlinecite{starikovskii2009sdbd}].

Spatial uniformity of the plasma is sought after for many SDBD applications. However, at sufficiently high discharge energy, the SDBD undergoes a transition in which the energy is concentrated into a few localized, high-current filaments~[\onlinecite{ding2019filamentary}]. This represents a major obstacle to expanding the range of plasma parameters possible for SDBD applications. For example, EAD thrust production is effective only when the plasma is uniformly generated lengthwise along a wingspan, and the transition to a filamentary regime limits the achievable thrust by SDBD~[\onlinecite{forte2007optimization,soloviev2011analytical,leonov2014dynamics}]. 

In previous work by Darny \textit{et al}~[\onlinecite{darny2020uniform}], the use of a barrier containing semiconducting material (silicon) was found to allow perfectly uniform propagation of the SIW, without streamers, at all times during plasma generation. For conventional SDBD in air at atmospheric pressure, streamers always formed during some phase of the discharge, typically during the positive-polarity phase when the SIW is cathode-directed~[\onlinecite{stepanyan2014nanosecond,bayoda2015nanosecond,zhu2017nanosecond}]. Furthermore, significantly higher current was reported compared to SDBD, which may point to stronger ionization. These observations were hypothesized to arise because photons generated by the gas-phase SIW absorb in silicon, causing photoexcitation of electron-hole pairs. In turn, these free charge carriers modify the electric field in the gas phase such that the seed electron avalanches ahead of the propagating SIW are initiated in closer proximity to the ionization front. The densification of seed electron avalanches promotes stability of the ionization front and suppresses the formation of streamers. This mechanism was proposed on the basis of a model showing how the branching of streamers depends on the stochasticity of photoionization~[\onlinecite{xiong2014branching}]. To support this hypothesis, a continuous-wave (cw) laser was used to irradiate the surface, for the purpose of photoexciting additional electron-hole pairs during plasma generation. As a result, the SIW deformed in the vicinity of the laser spot.

In this work, we aim to understand the temporal evolution of the laser-plasma interaction. We will refer to the discharge introduced by Darny \textit{et al}~[\onlinecite{darny2020uniform}] as the "semiconducting barrier discharge (SeBD)". Instead of employing cw irradiation, we will demonstrate the use of a pulsed laser and synchronized detection to characterize the nanosecond-scale response of the plasma. Fast imaging of optical emission and current-voltage measurements of the energy will reveal a key timescale for the interaction.

\section{\label{sec:level1}Experimental setup}
\begin{figure}[h]
    {\includegraphics[scale=0.65]{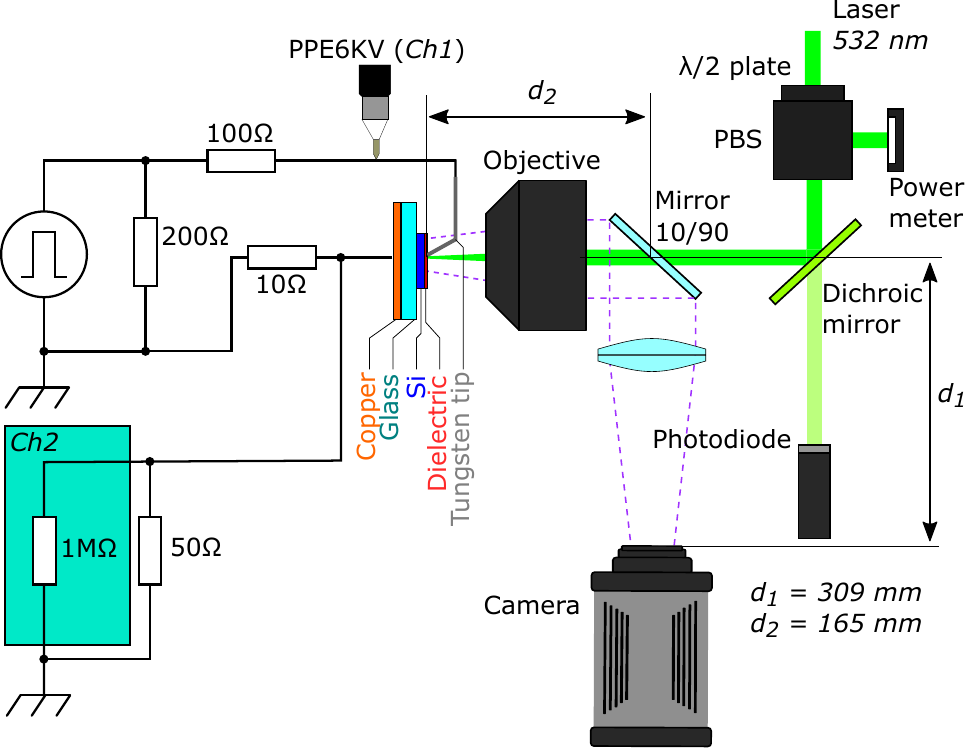}}
    \caption{\label{fig:experimental setup} Schematic diagram of the experimental setup for discharge generation, fast imaging, and laser irradiation. The oscilloscope is represented by its input channels labeled "Ch1" and "Ch2".}
\end{figure}

The SeBD discharge reactor and plasma conditions used in this work were identical to those presented by Darny \textit{et al}~[\onlinecite{darny2020uniform}] and will be briefly summarized here. As shown in Figure \ref{fig:experimental setup}, a tungsten wire electrode (\text{\diameter\ 100 µm}) was placed in mechanical contact with a wafer consisting of a 1-µm thick SiO$_2$ layer grown thermally on the polished side of a \textit{p}-type silicon substrate (1-20 $\Omega\cdot$cm$^{-1}$ resistivity, 525 µm thickness). To limit the current, the unpolished back side of this wafer was placed in contact with a 1-mm thick borosilicate glass plate, in turn covered on its back side by copper adhesive. A 10-$\Omega$ shunt resistor then connected the copper contact to ground. The SeBD was generated in open ambient air at atmospheric pressure.

For all the experiments presented in this work, high-voltage pulses 1.6 kV in amplitude and 30 ns in duration were applied to the tungsten wire at a pulse repetition frequency of 50 Hz to generate the SeBD. To minimize reflected power from the load, parallel and series resistors of 200 $\Omega$ and 100 $\Omega$, respectively, were inserted between the output coaxial cable of the high-voltage pulse generator and the reactor. The voltage applied to the tungsten wire electrode was measured using a passive probe (Lecroy PPE 6 kV) with 400-MHz bandwidth. The total current was determined by measuring the voltage across the 10-$\Omega$ shunt resistor using a 50-$\Omega$ coaxial cable, creating an effective current-sensing resistance of 8 $\Omega$. All current and voltage signals were acquired using an oscilloscope with 2-GHz bandwidth (Lecroy Waverunner 204 MXI).

The accuracy of the current-voltage measurements was verified by placing test impedances in place of the SeBD reactor, following the procedure used for similar measurements of nanosecond discharges~[\onlinecite{orriere2018ionization,pai2008etude}]. Using a test capacitance ($C$), current ($i(t)$) and voltage ($v(t)$) waveforms were confirmed to follow the relation $i = Cdv/dt$ accurately without any clear indication of deviations due to parasitic circuit elements. Likewise, with a test inductance ($L$) we determined that the current-voltage measurements followed the relation $v = Ldi/dt$ with similar accuracy. With this method, we measured the total current composed of the displacement current and conduction currents from the gas-phase discharge and also the silicon. The latter can originate from charge carrier generation and transport driven by the SeBD.

The optical system for laser irradiation of the wafer and fast imaging of the SIW are also shown in Figure \ref{fig:experimental setup}. The experimental setup was derived from a Raman spectrometer~[\onlinecite{pai_plasma-liquid_2021}] modified to provide the laser beam manipulation required in the present work. A diode-pumped solid-state laser (Elforlight Spot) emitted pulses 2 ns in duration at a wavelength of $\lambda=532$ nm. The laser power was adjusted by rotating a half-wave plate placed in front of a polarizing beamsplitter (PBS), both placed along the beam path ahead of reflection by a dichroic mirror (Semrock RazorEdge). Following reflection, the beam was soft-focused onto the wafer using a nominally 15× UV reflective microscope objective (Beck Optronic Solutions, model 5002) with a focal length of 13.4 mm and numerical aperture of 0.50. The working distance was adjusted to produce a spot size 65 µm in radius. Plasma emission was collected by the microscope objective, then partially reflected off a beamsplitter with 90\% transmission and 10\% reflection before focusing by an achromatic doublet UV lens (nominal 200 mm focal length) to form a sharply focused image of the discharge on the detector of an intensified CCD camera (Princeton Instruments PIMAX 4).

The imaging and laser irradiation parameters were as follows. By setting the distances $d_1 = 309$ mm and $d_2 = 165$ mm shown in Figure \ref{fig:experimental setup}, the magnification of the system projected 1.52×1.52 µm$^2$ from the object plane onto each pixel of the CCD detector. Plasma imaging was performed in single-shot mode with an exposure time (gate width) of 3 ns. The gate delay of the camera was varied to acquire images at different times during discharge evolution. The laser power irradiating the wafer was measured by replacing the SeBD reactor with a power meter (Thorlabs S121C) placed close to the object plane of the microscope objective. For all the experiments presented in this work, the average laser power at the target was measured to be 8.5 µW, corresponding to an energy per laser pulse of 0.17 µJ at the repetition frequency of 50 Hz. Given the spot size, this amounts to a fluence of $F=1.3$ mJ/cm$^2$ per pulse at the wafer surface.

To monitor the stability of the laser power during SeBD irradiation experiments, the power meter was repositioned to measure the part of the beam reflected off the polarizing beamsplitter, as illustrated by Figure \ref{fig:experimental setup}. The power at this location was measured to be $68\pm17$ µW, implying a $\pm25$\% fluctuation at the wafer surface that did not cause any observable variation in the intensity of plasma emission. Over the course of the experiments, the average power drifted by $\pm0.1$\%. 

The high-voltage pulse generator, laser, and camera were synchronized using a delay generator (Stanford Research Systems DG645), with a repetition frequency of 50 Hz for all devices. A photodiode with a nominal rise time of 150 ps (Thorlabs DET025A/M) was placed behind the dichroic mirror to measure the arrival time of the laser pulse at the wafer, after taking into account the difference in free-space delay between the separate optical paths to the wafer and photodiode. The effect of irradiation was studied by varying the delay of the laser relative to the plasma.

\section{\label{sec:level1}Results and discussion}
\begin{figure}[h]

    \includegraphics[width = 0.4\textwidth]{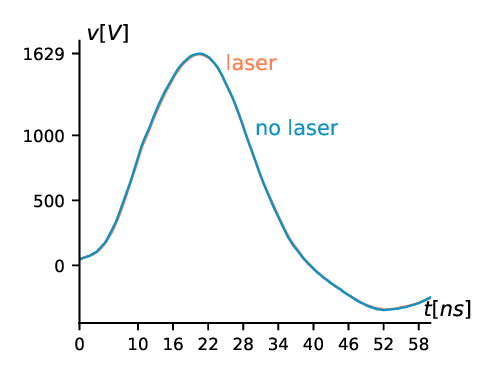}
    \vfill
	\includegraphics[width = 0.4\textwidth]{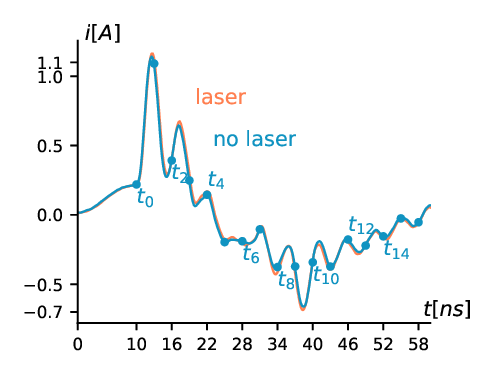}
	\vfill
	\includegraphics[width = 0.4\textwidth]{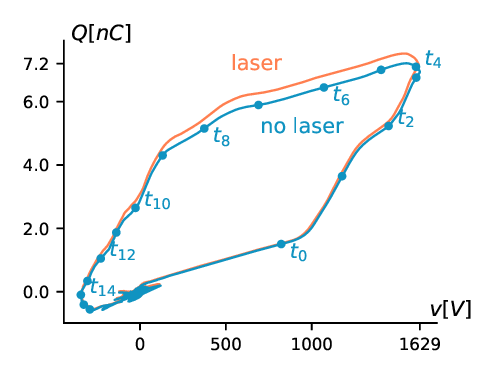}
     
	\caption{Applied voltage (top) and total current (middle) waveforms of the SeBD, without (blue) and with (orange) 2-ns pulsed laser irradiation starting at $t = -35$ ns, covering the region on the wafer surface indicated in Figure \ref{fig:plasma images}. Also shown is the corresponding charge-voltage Lissajous plot (bottom). Points indicate the different start times of camera gating.}
	\label{fig:Lissajous curves}
\end{figure}

The imaging and current-voltage measurements of the SeBD without laser irradiation were presented in detail by Darny \textit{et al}~[\onlinecite{darny2020uniform}]. Nonetheless, we repeated these characterizations here because we improved the accuracy of measuring current and voltage. In this way, a comparison can be made between measurements with and without irradiation, as shown by the example in Figure \ref{fig:Lissajous curves}(top, middle). Typically, an increase in current amplitude is measurable but small, at most a few percent. However, an increase in energy is more apparent upon plotting the charge-voltage Lissajous curve (Figure \ref{fig:Lissajous curves}(bottom)), where the charge $Q(t)$ is found by integrating the measured $i(t)$ in time. The Lissajous figure closes well in a loop, demonstrating that minimal memory charge remains after each discharge.

\begin{figure*}[t!]
		\centering
            \includegraphics[width = 0.15\textwidth]{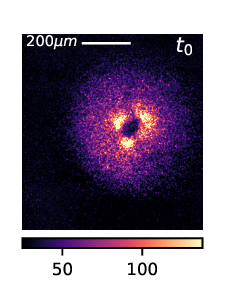}
		\includegraphics[width = 0.15\textwidth]{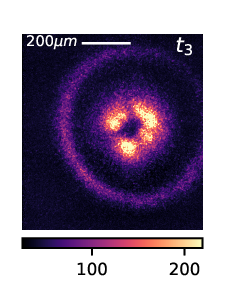}
		\includegraphics[width = 0.15\textwidth]{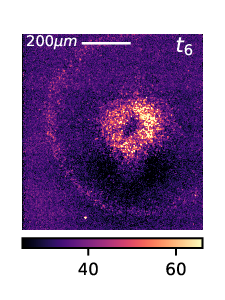}
		\includegraphics[width = 0.15\textwidth]{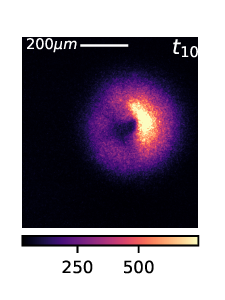}
		\includegraphics[width = 0.15\textwidth]{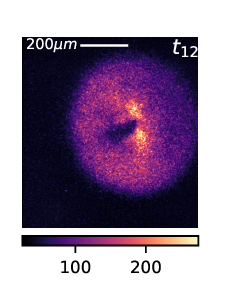}
		\includegraphics[width = 0.15\textwidth]{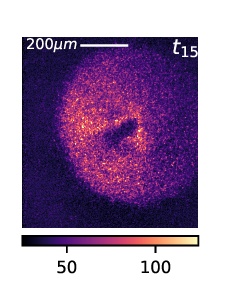}
		\includegraphics[width = 0.15\textwidth]{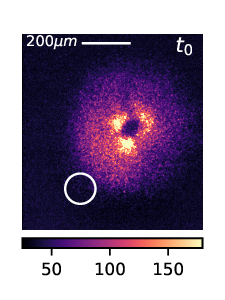}
		\includegraphics[width = 0.15\textwidth]{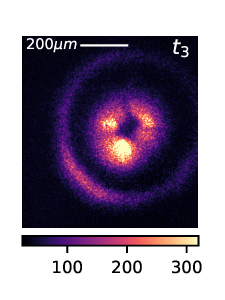}
		\includegraphics[width = 0.15\textwidth]{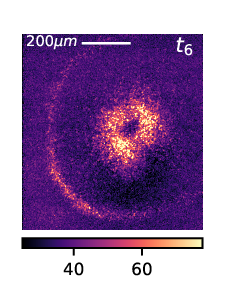}
		\includegraphics[width = 0.15\textwidth]{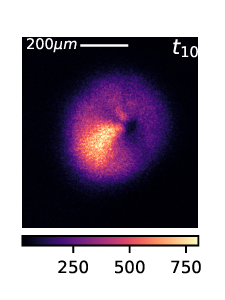}
		\includegraphics[width = 0.15\textwidth]{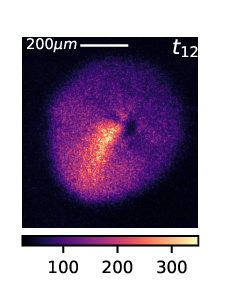}
		\includegraphics[width = 0.15\textwidth]{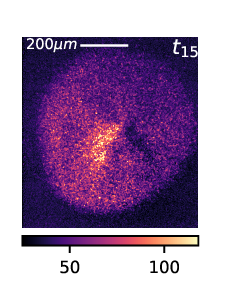}
		\caption{Single-shot images of the SeBD at different times without (top) and with (bottom) laser irradiation starting at $t = -35$ ns. The region of irradiation is indicated (circle) in the bottom image for $t_0$.}
		\label{fig:plasma images}
	\end{figure*}

Figures \ref{fig:Lissajous curves}(bottom) and \ref{fig:plasma images}(top) respectively follow the energy dissipation and SIW propagation as time progresses, for the SeBD without laser irradiation. The slope of the Lissajous figure prior to breakdown at $t_0 = 10$ ns represents the static capacitance of the reactor without plasma. From $t_0$ to $t_2 = 16$ ns, breakdown commences with the appearance of the initial and highest current peak. Also, the slope of the Lissajous curve increases, indicating an increase in capacitance due to the generation of the SeBD. Simultaneously, a localized high-intensity corona forms near the tungsten wire electrode, and the rest of the discharge forms a homogeneous disk. The SIW first appears at $t_2$, at the end of the initial current peak, taking on a circular ring shape, as shown in Figure \ref{fig:plasma images}(top) at $t_3 = 19$ ns. From $t_2$ to $t_4 = 22$ ns, a second current peak appears as the voltage reaches its maximum value, and the SIW propagates outward and maintains a high light intensity. The SIW fades and leaves only the corona from $t_4$ to $t_9 = 37$ ns, as shown in Figure \ref{fig:plasma images}(top) at $t_6 = 28$ ns. The slope of the Lissajous curve returns to its pre-discharge value, indicating that the plasma-induced component of the capacitance has been switched off. The negative-phase SIW begins to expand noticeably starting at $t_{10} = 40$ ns and continuing until $t_{15} = 55$ ns, coinciding with a change in the slope of the Lissajous figure indicating the return of a plasma-induced capacitance.

The effect of laser irradiation on the energy dissipation and SIW propagation of the SeBD is shown in Figures \ref{fig:Lissajous curves}(bottom) and \ref{fig:plasma images}(bottom), respectively, for one example case where irradiation occurred over the region indicated in Figure \ref{fig:plasma images}(bottom) and was timed to begin at $t = -35$ ns, before the arrival of the SIW at this location. In this particular example, the total energy determined by the area enclosed by the Lissajous curve increases from 6.4 µJ without laser irradiation to 6.9 µJ with irradiation. Even though the laser pulse duration is only 2 ns, the effect on the plasma lasts throughout the discharge duration. This is apparent from the sequence of images in Figure \ref{fig:plasma images}, where the SIW is clearly more intense and propagates further towards the zone of laser irradiation during the positive phase of the discharge. During the negative phase, the SIW exhibits a similar protrusion or bulging in this region.
	
\begin{figure}[t!]
		\includegraphics[width = 0.5\textwidth]{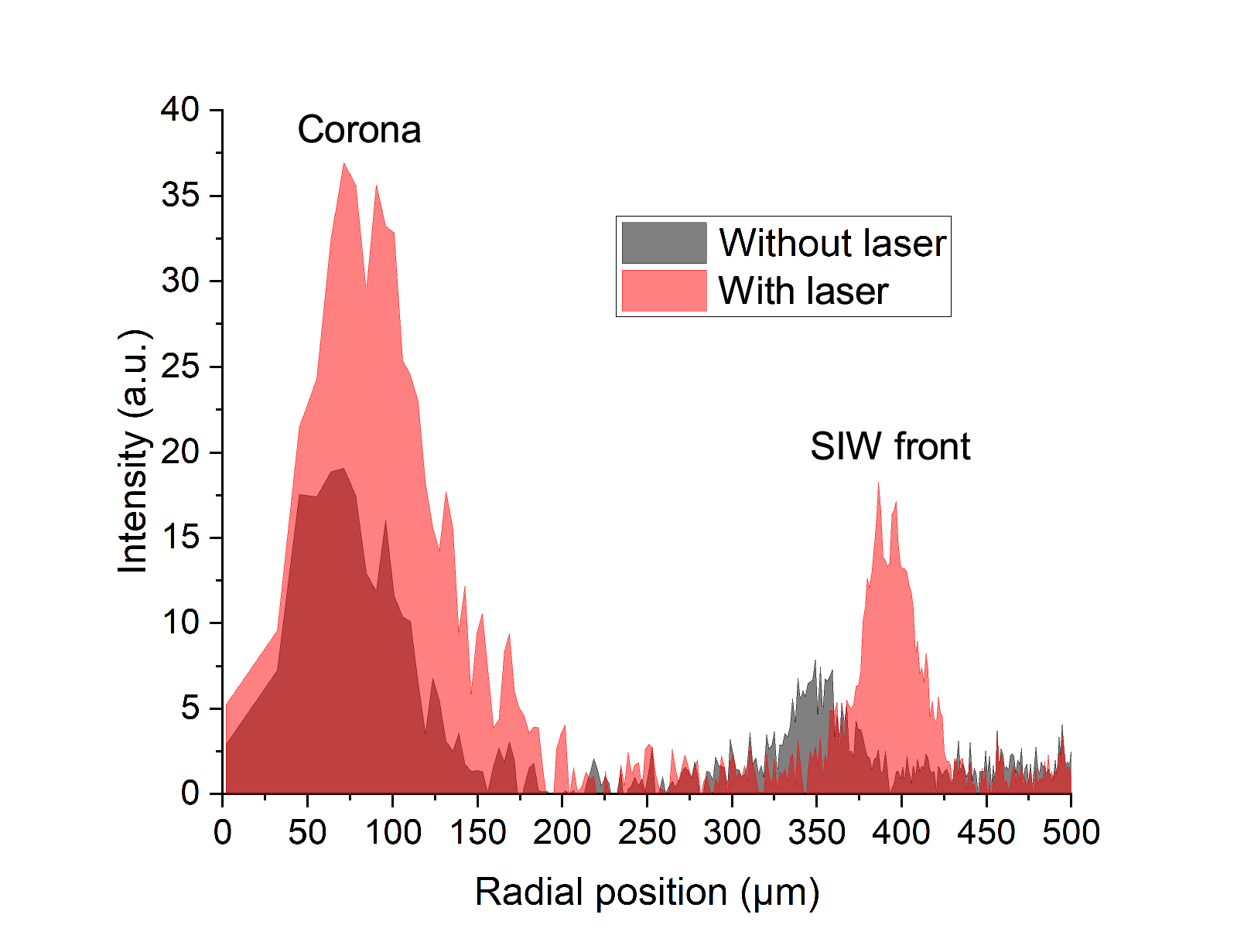}
		\caption{Radial profiles of optical emission intensity integrated over the angular range $\Delta\theta$ from $\theta = -119$° to $\theta = -145$°, without and with laser irradiation starting at $t = -35$ ns. The 3-ns exposure time begins at $t_6=$ 28 ns. Baseline subtraction was applied to permit fitting of the SIW front.}
		\label{fig:radial profiles}
	\end{figure}

To perform quantitative image analysis, we integrate the detected emission over the angular range $\Delta\theta$ from $\theta = -119$° to $\theta = -145$° where the plasma deformation is apparent from Figure \ref{fig:plasma images}. The data were sorted according to the radial position and smoothed by 100-point adjacent averaging. The result is a radial profile of the data points within $\Delta\theta$, as shown in Figure \ref{fig:radial profiles} for the same time $t_6=28$ ns as the two images in Figure \ref{fig:plasma images}. The corona and SIW front are clearly identifiable. The SIW front was fitted by the sum of two Gaussian functions, and the radial position of the peak of the fitting function defines the position of the SIW front. Without irradiation, at $t_6$ the corona region extends up to $r=100$ µm, and the SIW front is positioned at $r=230$ µm. With irradiation, at $t_6$ the corona expands to $r$ = 120 µm, while the SIW front propagates further to $r$ = 260 µm. Also, both the corona and SIW front increase in total emission intensity by 89\% and 82\%, respectively.

Figure \ref{fig:center,intensity,energy)}(top) shows the position of the SIW front as a function of time for laser irradiation occurring at different times $t$ corresponding to delays $\tau_d = -t$. Before the SeBD experiences irradiation, the SIW front propagates from $r = 243$ µm at $t = 16$ ns to $r = 351$ µm at $t = 28$ ns. Its velocity, calculated as the slope between consecutive data points, decreases over the course of this displacement from \num{1.6e6} to \num{1.5e5} cm/s, similar to the values found by Darny \textit{et al}. Immediately after halting irradiation of the SeBD, the SIW front propagates less far than before the initial irradiation experiment. The cause of this variation can be attributed to a memory effect observed by Darny \textit{et al}, where an absence of plasma appeared in the location of the irradiation spot after cw laser irradiation was switched off. This void persisted for at least several minutes, pointing to a yet unidentified process with a very long lifetime. Darny \textit{et al} suggested that this could reflect the influence of long-lived trapped charges at the Si-SiO$_2$ interface. 

With irradiation, no difference in propagation is noticeable for irradiation delays $\tau_d = -t =  3235-5235$ ns, which is well before the discharge. At $\tau_d = 2235-2735$ ns, the SIW front begins to propagate farther and faster, though within the variation of the measurements without irradiation. For $\tau_d \leq 1235$ ns, the behavior becomes nearly independent of $\tau_d$, with the SIW front initiating at $t = 16$ ns with its center at $r = 253$ µm and a velocity of \num{1.8e6} cm/s, which is similar to the initial velocity without irradiation. However, the SIW front propagates farther with irradiation, reaching $r = 403$ µm at $t = 31$ ns due to less deceleration over the course of travel.

Also shown in Figure \ref{fig:center,intensity,energy)}(center) is the corresponding optical emission intensity of the SIW integrated over $\Delta\theta$ and the radial range corresponding to the full-width at half-maximum of the two-Gaussian fitting function. Before the SeBD experiences irradiation, the intensity of the SIW front diminishes with time down to $5\%$ of its initial value by $t=28$ ns, at a rate of change significantly faster than those of the surface areas of the SIW front at $\propto 1/r$ or the SeBD as a whole at $\propto 1/r^{2}$. Immediately after halting irradiation of the SeBD, the intensity decays with time at a rate faster than beforehand.

With irradiation, the intensity of the SIW front decays with time in a similar manner, but the curves shift to higher intensity. From $\tau_d = 5235$ to 2235 ns, the intensity curves shift upwards progressively but still within the variation of the measurements without irradiation. For $\tau_d \leq 1235$ ns, the curves shift to even higher intensity and nearly overlap each other. Thus, the increase in emission intensity closely follows the increase in the SIW front position shown in Figure \ref{fig:center,intensity,energy)}(top).

Finally, Figure \ref{fig:center,intensity,energy)}(bottom) shows corresponding measurements of the total energy of the SeBD. Laser irradiation without plasma alone does not result in any measurable energy deposition. Without irradiation but with plasma, the energy is 6.45±0.11 µJ per pulse. From $\tau_d=-t = 5235$ to 3235 ns, irradiation has no effect on the energy. However, the variation of energy measurements at a given time delay begins to shift upwards starting at $\tau_d = 2735$ ns. The increase in total energy rises completely above uncertainty at $\tau_d = 1235$ ns, the same delay at which the SIW front position and emission intensity also become unambiguously greater than their respective values without irradiation. Overall, the energy increases linearly with $t$ starting at about $\tau_d = 3$ µs, reaching 6.9±0.1 µJ per pulse for $\tau_d = 35$ ns, corresponding to irradiation just prior to discharge generation.

    \begin{figure}[t!]
		\includegraphics[width = 0.4\textwidth]{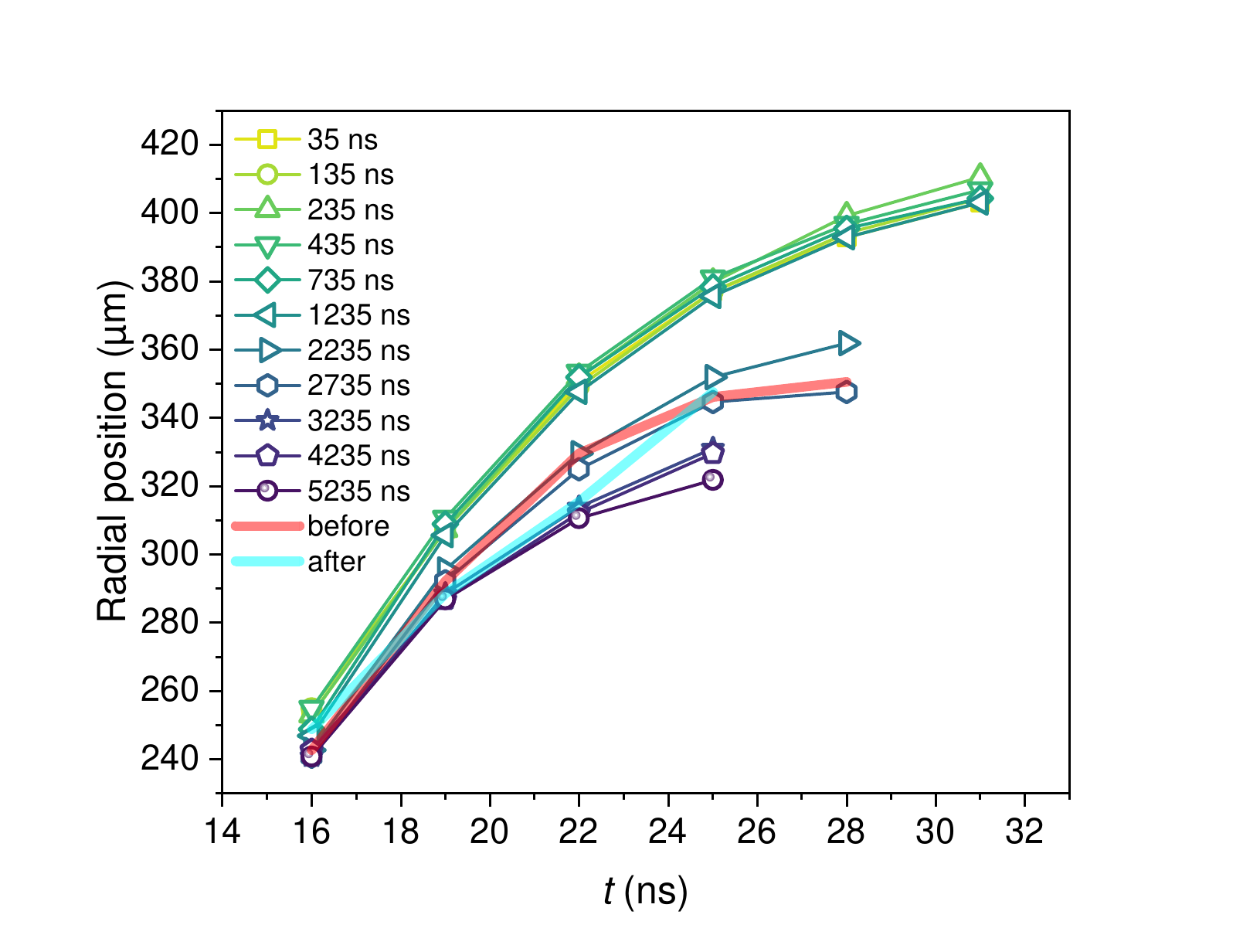}
		\vfill
		\includegraphics[width = 0.4\textwidth]{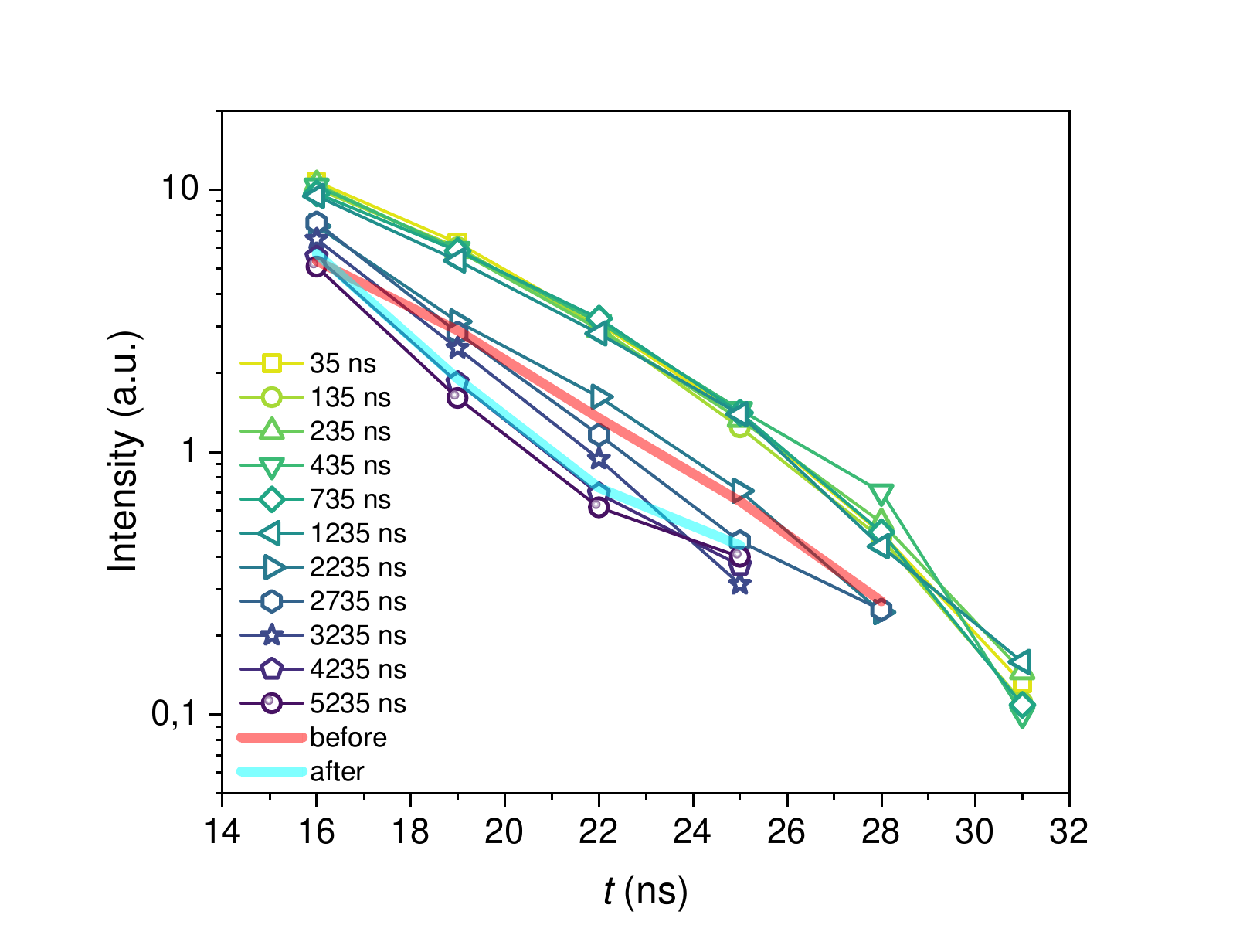}
		\vfill		
		\includegraphics[width = 0.4\textwidth]{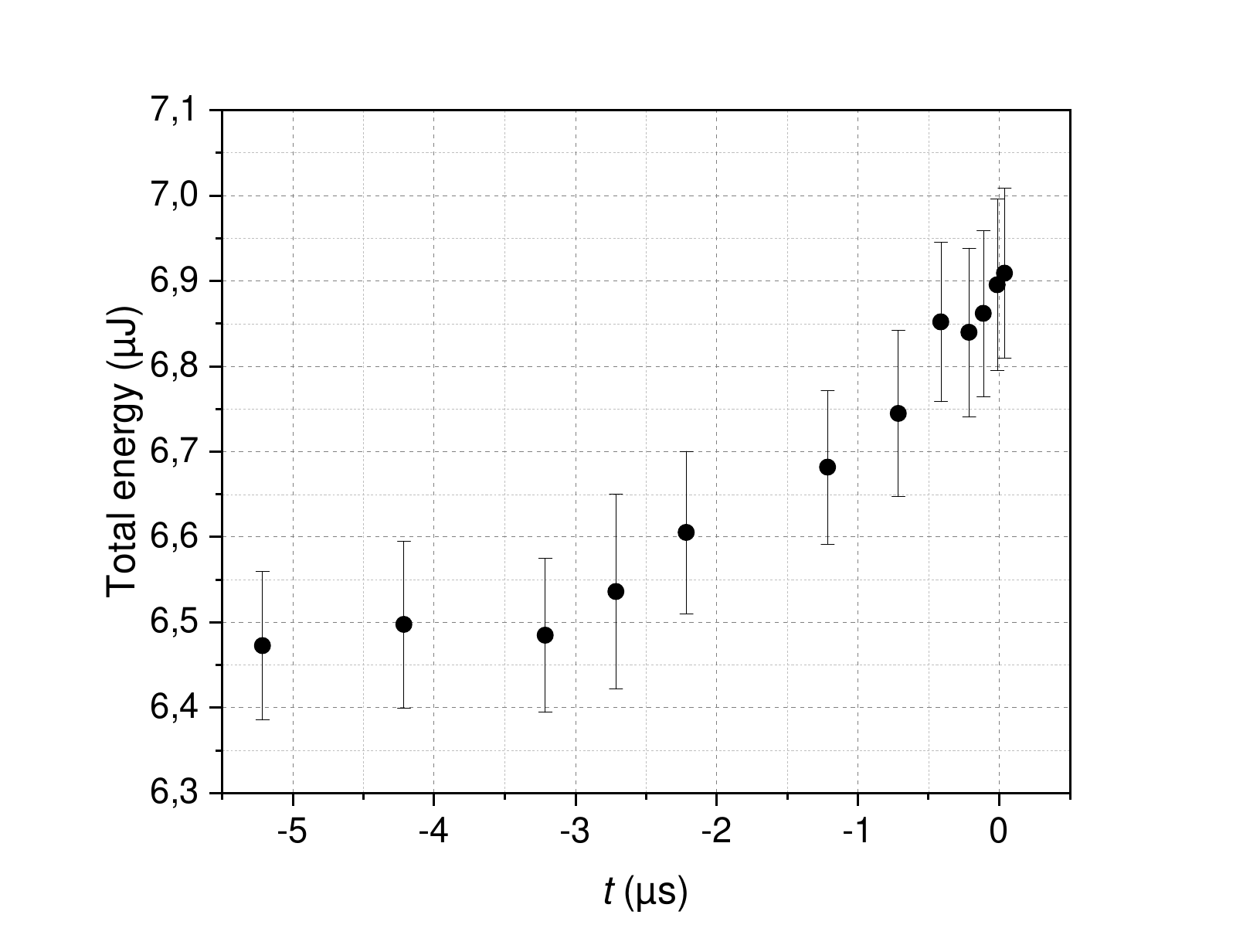}
		\caption{Radial position (top) and integrated optical emission intensity over the angular range $\Delta\theta$ (middle) of the SIW front, as well as total energy of the SeBD (bottom) as a function of the time delay $\tau_d=-t$ of the pulsed laser. Also shown are measurements taken before the SeBD experiences irradiation and immediately after switching off the laser.}
		\label{fig:center,intensity,energy)}
	\end{figure}

The above results demonstrate that the cause of the SIW stimulation is not related to the liberation of surface charge by laser irradiation. Charges deposited by plasmas are held in surface trap states with energies on the order of 1 eV~[\onlinecite{guaitella2011role}]. Thus, the photon energy used in this study is sufficient to free the trapped charge~[\onlinecite{ussenov2024laser}]. Desorption would then eliminate screening of the applied electric field by surface charge deposited by previous discharges. However, this should not depend on the time delay before SeBD generation, and therefore the desorption of surface charge does not explain the results shown in Figure \ref{fig:center,intensity,energy)}.

The fact that irradiation does not have to coincide with plasma generation to produce a stimulation of the SeBD also implies that the mechanism of interaction is different from the operating principles of the hybrid microplasma/semiconductor photodetector~[\onlinecite{park2002photodetection,ostrom2005microcavity}] and the plasma bipolar junction transistor~[\onlinecite{wagner2010coupling,tchertchian2011control}]. Like the SeBD, these silicon-based devices rely on the photoexcitation of electron-hole pairs upon irradiation by the gas-phase plasma and/or an external light source. The electrons then tunnel through the plasma-semiconductor interface with the help of electronic band-bending induced by the electric field of the plasma sheath, in a secondary electron emission process~[\onlinecite{li2013modulating}] leading to electron multiplication in the gas phase. However, in the case of the SeBD, there is no electric field at the times of laser irradiation $t=-\tau_d$ presented in Figure \ref{fig:center,intensity,energy)}. Moreover, the 1-µm thick SiO$_2$ layer prevents electron tunneling from the silicon into the gas phase. Previous research on microplasmas confined within silicon microcavities similarly featured a dielectric layer~[\onlinecite{park200540000pixel}], but no evidence of SIW propagation was presented nor was the effect of external irradiation investigated. 

Furthermore, a comparison of the extra current resulting from irradiation indicates that the nature of plasma-semiconductor coupling for the SeBD should be distinct from that for the microplasma-based photodetector or transistor. Irradiation of the SeBD at $t=-35$ ns increases its charge by $\delta Q\sim0.4$ nC per pulse (Figure \ref{fig:Lissajous curves}), and given the fluence $F=1.3$ mJ/cm² per pulse, the additional current generated per unit power density of irradiation is $\delta Q/F\sim 0.3$ µA/W/cm². This is much lower than $\sim 5000$ A/W/cm² for the microplasma photodetector, calculated based on its photosensitivity of $\sim 0.5$ A/W at a wavelength of 532 nm and a microcavity area of $100\times100$ µm²~[\onlinecite{ostrom2005microcavity}]. The considerable difference in this figure of merit points to different coupling mechanisms.

Before applying the high-voltage pulse, carrier transport within the silicon is limited to diffusion. Given the absorption length of 1 µm in silicon at $\lambda=532$ nm and the size of the irradiation spot, the excess carrier density generated by the laser should be $\Delta n\sim10^{19}$ cm$^{-3}$ assuming a quantum efficiency of $\sim1$, placing the electron-hole plasma in the high-injection (HI) regime. For a thermal oxide \textit{p}-doped wafer with a doping level similar to this study, also irradiated by a pulsed laser at 532 nm into the HI regime, time-resolved measurements showed that the excess carrier concentration remained in HI for several tens of µs during the recombination phase~[\onlinecite{grivickas2000spatially}]. Under these conditions, transport is driven by ambipolar diffusion, with a constant of $D_a=2.5$ cm²/s measured for $\Delta n\sim$ \num{4e19} cm$^{-3}$~[\onlinecite{mouskeftaras2016direct}]. The value of $D_a$ should increase with decreasing $\Delta n$~[\onlinecite{rosling1994ambipolar}]. Therefore, the minimum diffusion length corresponding to the critical delay $\tau_d = 3$ µs is $L_d=\sqrt{D_a\tau_d}\sim 30$ µm. This suggests that for $\tau_d > 3$ µs, the photoexcited carriers diffuse far from the Si-SiO$_2$ interface, limiting the ability of these charges to stimulate the SeBD. In particular, $L_d$ is much larger than the size of the depletion layer at the Si-SiO$_2$ interface ($\sim1$ µm under strong inversion conditions) expected to form during plasma generation, proposed by Taihi \textit{et al}~[\onlinecite{taihi2026photonic}] as the region responsible for driving the stimulation of the SeBD by irradiation at intensities lower than employed in this study.

In conclusion, pulsed laser irradiation of the SeBD surface has revealed that stimulation of the SIW front and total energy occur only when irradiation occurs within about 3 µs before plasma generation. This timescale excludes surface charge desorption by irradiation as the cause of the stimulation, pointing instead to the relevance of the diffusion of photoexcited carriers away from the Si-SiO$_2$ interface. These results will inform future research into the precise mechanism of SIW propagation of the SeBD, which could potentially be an interesting alternative to DBDs for a broad range of applications~[\onlinecite{brandenburg2023barrier}]. The SeBD can also be envisioned as a potential foundation for plasma-based optoelectronic devices capable of operating in open ambient air.

% If in two-column mode, this environment will change to single-column format so that long equations can be displayed. 
%\begin{widetext}
%$$\mbox{put long equation here}$$
%\end{widetext}

% Figures should be put into the text as floats. 
% Use the graphics or graphicx packages (distributed with LaTeX2e).
% Put the label that you will use with \ref{} command in the braces of the \label{} command.
%
% \begin{figure}
% \end{figure}

% \begin{table}
% \end{tabular}
% \end{table}

\begin{acknowledgments}
We gratefully acknowledge financial support from the Agence Nationale de la Recherche program JCJC PLASMAFACE (ANR-15-CE06-0007-01) and the ‘Investissements d’Avenir’ program LABEX INTERACTIFS (ANR-11-LABX-0017-01) of the French government, as well as the CPER-FEDER program of the Région
Nouvelle Aquitaine.
\end{acknowledgments}

\bibliography{aipsamp}

\end{document}